\documentclass[twocolumn]{aastex62}
\usepackage{hyperref}

\usepackage{subfigure}

\newcommand{\ssymbol}[1]{^{\@fnsymbol{#1}}}

\def\beq{\begin{equation}}
\def\eeq{\end{equation}}
\def\bi{\begin{itemize}}
\def\ei{\end{itemize}}
\def\ben{\begin{enumerate}}
\def\een{\end{enumerate}}
\def\bea{\begin{eqnarray}}
\def\eea{\end{eqnarray}}

\definecolor{lightblue}{rgb}{0.19, 0.55, 0.91}
\newcommand{\cm}{{~{\color{black}}}~}

\newcommand{\gaia}{{\it Gaia}\ }
\usepackage[all]{hypcap}
\defcitealias{iptadr2}{IPTA DR2}
\defcitealias{dcl+16}{D16}
\defcitealias{cl02}{NE2001}
\defcitealias{ymw+17}{YMW16}
\defcitealias{dvt+08}{D08}
\defcitealias{kkn+16}{K16}

\shorttitle{Improving pulsar distances with \gaia}
\shortauthors{Mingarelli et al.}

\begin{document}

\title{Improving Binary Millisecond Pulsar Distances with Gaia} 

\correspondingauthor{Chiara M. F. Mingarelli}
\email{chiara.mingarelli@uconn.edu}

\author[0000-0002-4307-1322]{Chiara M. F. Mingarelli}
\affil{Department of Physics, University of Connecticut, 196 Auditorium Road, U-3046, Storrs, CT 06269-3046, USA}
\affil{Center for Computational Astrophysics, Flatiron Institute, 162 Fifth Ave, New York, NY 10010, USA}

\author[0000-0001-5725-9329]{Lauren Anderson}
\affil{Center for Computational Astrophysics, Flatiron Institute, 162 Fifth Ave, New York, NY 10010, USA}
\affil{Observatories of the Carnegie Institution for Science, 813 Santa Barbara Street, Pasadena, CA 91101, USA}

\author[0000-0001-9907-7742]{Megan Bedell}
\affil{Center for Computational Astrophysics, Flatiron Institute, 162 Fifth Ave, New York, NY 10010, USA}

\author[0000-0002-5151-0006]{David N. Spergel}
\affil{Center for Computational Astrophysics, Flatiron Institute, 162 Fifth Ave, New York, NY 10010, USA}
\affil{Department of Astrophysical Sciences, Princeton University, Peyton Hall, Princeton, NJ 08544-0010, USA}

\author[0000-0002-6437-5229]{Abigail Moran}
\affil{Department of Physics, University of Connecticut, 196 Auditorium Road, U-3046, Storrs, CT 06269-3046, USA}

\begin{abstract}

Improved distance measurements to millisecond pulsars can enhance pulsar timing array (PTA) sensitivity to gravitational-waves, improve tests of general relativity with binary pulsars, improve constraints on fuzzy dark matter, and more. Here we report the parallax and distance measurements to seven \gaia DR2 objects associated with seven International PTA pulsars: J0437-4715, J1012+5307, J1024-0719, J1732-5049, J1910+1256, J1843-1113, and J1949+3106. 
By multiplying the posteriors of the PTA-based parallax measurements with the \gaia parallax measurement to the pulsar's companion, we improve the distance measurements from a few percent to a factor of five, and a tentative detection of a binary companion to J1843-1113. We also find an order of magnitude improvement in the parallax measurement to J1949+3106.
\end{abstract}



\section{Introduction}
\label{sec:intro}
Millisecond pulsars (MSPs) are the best clocks in nature~\citep{bkh+82}, and as such, are used in a wide variety of physics experiments ranging from test of dark matter (e.g. \citealt{pzl+18}) to low-frequency gravitational-wave detection \citep{saz78, det79, hd83} and test of General Relativity~\citep{lwj+09, stw+11, lyp+18}.

Precise pulsar distance measurements are particularly important for pulsar timing array (PTA) experiments~\citep{ng12p5, dcl+16, mhb+13}, which aim to detect nanohertz gravitational waves from both individual inspiralling supermassive black hole binaries \citep{babak+16, cgw19} and the gravitational wave background from their cosmic merger history \citep{p01, ng12p5, M19}. In particular, precise pulsar distance measurements can improve the sensitivity of PTAs to continuous gravitational wave sources~\citep{zwx+16}, and make it possible to measure the evolution of supermassive black hole binaries over thousands of years~\citep{mgs+12}. 

Direct pulsar distance measurements are obtained either from pulsar timing parallax measurements -- the curvature of the wavefront will cause a variation in pulse arrival times at different positions throughout the year, e.g. \cite{lk04}-- or in the rare case that the pulsar is relatively nearby, via direct Very Long Baseline Interferometry (VLBI) measurements, e.g. ~\cite{dvt+08}. Distances can also be independently inferred from measuring the binary orbital period ($P_b$) derivative when possible, via $\dot{P}_b/P_b\propto D_K$ (see e.g. \citealt{s70}). This kind of distance measurement is called a kinematic distance, which we denote by $D_K$, \cm{and by annual orbital parallax measurements (e.g. \citealt{Kopeikin}).}

When direct distance measurements are not available, distances to MSPs are estimated via their Dispersion Measure (DM) -- a frequency-dependent delay in the pulse arrival time due to the radio waves traveling through the ionized interstellar medium of the galaxy, see e.g.~\citep{pulsarAstronomy}. This delay is proportional to the product of the mean electron number density, $ n_e$, and the distance to the pulsar, $D$. This distance estimate is limited by the uncertainties in the galactic
electron distribution. Such models have been proposed by~\cite{cl02}, from hereon \citetalias{cl02}, and more recently by \cite{ymw+17}, henceforth 
\citetalias{ymw+17}.
However, since the relationship between DM and pulsar distances can only be calibrated by pulsars with well-known distances from e.g. timing parallax, there are large errors associated with these inferred distance measurements, of the order of 20-40\%.

Due to their formation history, MSPs are often in binary systems \citep{bvdh91, tvdh06} with companions including other neutron stars, white dwarfs, and main sequence stars \citep{vdh84}. While distance measurements to the pulsars can be challenging, their binary companions offer a new avenue to measure the binary's distance. Indeed, optical data from the \gaia mission \citep{gaiadr2} can provide position, proper motion, and parallax measurements to some of these companions. This method is also not without its challenges, as the white dwarfs are typically dim at kpc distances, where most MSPs are found, and thus have a low signal-to-noise (S/N) measurement. In the future, NASA's Nancy Grace Roman Space Telescope (Roman, formerly WFIRST; \citealt{WFIRST}) will also be able to provide accurate astrometric measurements to much fainter magnitudes so that future analyses will be able to obtain accurate distances to many more pulsars.

Here we present a general framework to cross-match millisecond pulsars in the International Pulsar Timing Array Data,  \citep[][hereafter \citetalias{iptadr2}]{iptadr2}, and \gaia Data~(DR2; \citealt{gaiadr2}), in an effort to identify binary companions to the MSPs, and provide a new parallax (and therefore distance) measurement to the binary systems. \cm{We will continue to update our results with subsequent data releases, e.g. \cite{GaiaEDR3}. }

While \cite{JenningsEtAl:2018} report parallaxes and proper motions for cross-matches to known binary pulsars in the ATNF pulsar catalog, we restrict ourselves to the careful analysis of the subset of pulsars in \citetalias{iptadr2}, as these will have the greatest impact on the upcoming detection of nanohertz gravitational waves.

We find parallax measurements to seven binary MSPs: J0437-4715, J1012+5307, J1024-0719, J1732-5049, J1910+1256, J1843-1113, and J1949+3106. \cite{JenningsEtAl:2018} identified WD companions for J0437-4715, J1012+5307, J1024-0719, but missed J1732-5049, J1910+1256, J1843-1113, and J1949+3106 which we report here. We then go further and multiply the normalized parallax posteriors from the different experiments and get an improved distance measurement. 

We also tentatively identify a binary companion to J1843-1113, which we will monitor with new \gaia data releases, report an order of magnitude improvement in the parallax measurement to J1949+3106 with \gaia, and give the first parallax measurement to J1732-5049.

Every result in this work is reproducible via our public software, written in Python, and available on github, \url{https://github.com/ChiaraMingarelli/gaia_pulsars}. 

\section{Identifying binary candidates in \gaia}

\begin{figure*}[ht]
\centering     
\subfigure[Parallax]
        {
		\includegraphics{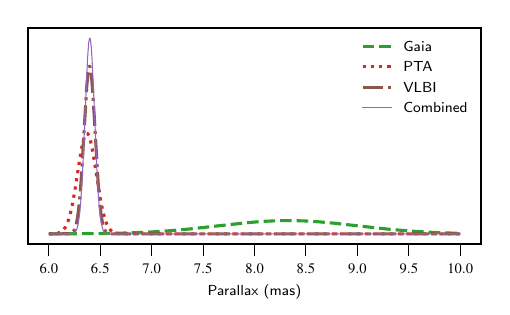}
        }
\subfigure[Distance]
        {
		\includegraphics{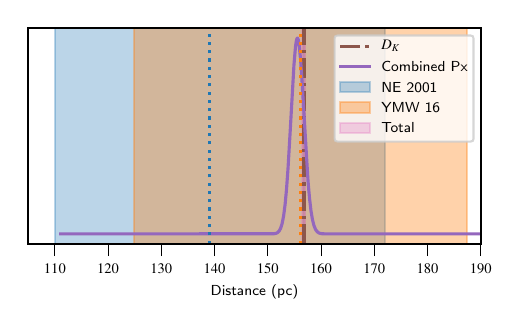}
        } \\
\subfigure[\gaia and IPTA RA]
        {
		\includegraphics{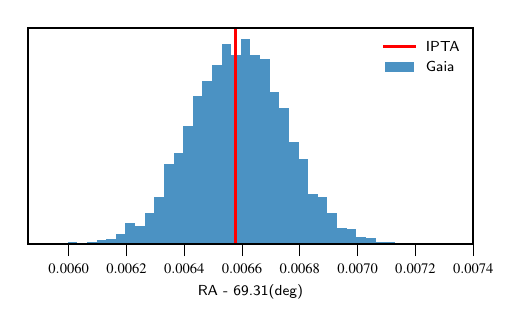} 
        }   
\subfigure[\gaia and IPTA Dec]
        {
		\includegraphics{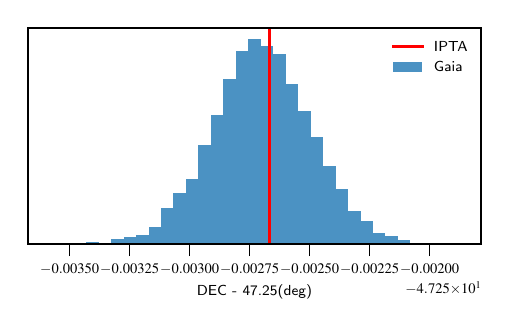} 
        }
\caption{Comparison of IPTA and \gaia values for J0437-4715. Clockwise from top right: (a) green dashed lines are \gaia parallax measurements to the binary system, red dashed-dot curves are based on pulsar timing distance measurements, and the solid purple curve is the final combined parallax. DM-based distance estimates from \citetalias{ymw+17} are shown in the yellow strip and \citetalias{cl02} are blue, with their mean values as a dashed line. These DM distances are illustrative only, and are not used in any distance calculation here. In (b) the resulting parallax measurement from (a) can be further combined with a dynamical distance estimate, $D_K$, to further improve the distance measurement. \cm{We show in (c) and (d) that the RA and DEC of the \gaia object is consistent with the IPTA-reported measurement (errors on IPTA sky location are too small to be seen). Proper motions also need to be within $3\sigma$ of PTA values for the cross-match to be accepted. }}
\label{fig:0437}
\end{figure*}

\begin{figure*}[ht!]
\centering     
\subfigure[Parallax]
        {
		\includegraphics{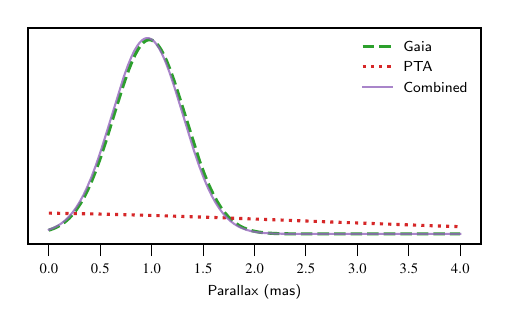}
        }
\subfigure[Distance]
        {
		\includegraphics{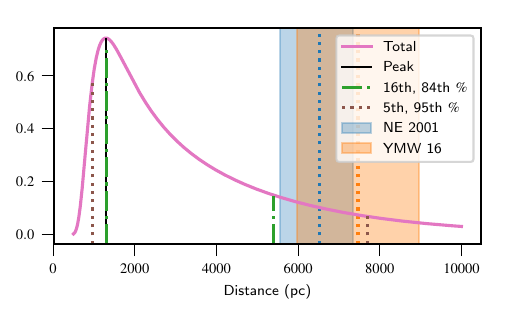}
        } \\
\subfigure[\gaia and IPTA RA]
        {
		\includegraphics{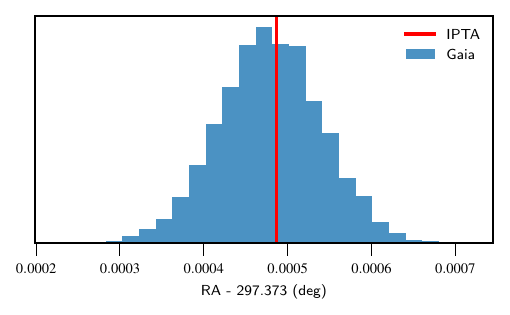} 
        }   
\subfigure[\gaia and IPTA Dec]
        {
		\includegraphics{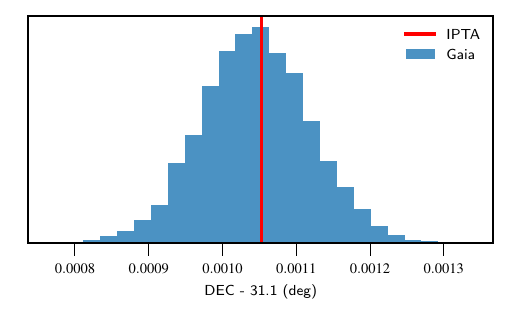} 
        }
\caption{Comparison of IPTA and \gaia values for J1949+3106, \cm{highlighting the usefulness of our approach.} Clockwise from top right as for the other pulsars. We show the details for this system in contrast to J0437-4715: here the PTA data are quite poor and truly benefit from the addition of \gaia data, as previous pulsar timing parallax measurements were negative. We also note that the both the YM16 and NE2001 galactic electron density models overestimate the distance to this source by approximately 5~kpc.}
\label{fig:1949}
\end{figure*}

Using \citetalias{iptadr2} we cross-referenced the MSP locations with objects in \gaia DR2~\citep{gaiadr2}, updating the pulsar positions to the \gaia epoch, and searching around the pulsar position within $3''$. The resulting cross-match returns 18 objects, to which we apply the following cuts: we require that the object has a parallax measurement \cm{which appears} in \gaia DR2, and that the \gaia proper motion in both RA and Dec are within $3\sigma$ of the \citetalias{iptadr2} value. \cm{We show a few examples of the resulting RA and DEC values which survive this cut in Figure \ref{fig:0437} and Figure \ref{fig:1949}}. We do not cut on absolute magnitude, since some pulsars, such as J1024-0719, have main sequence stars as binary companions (\autoref{fig:HRD}). Moreover, we search through all pulsars in \citetalias{iptadr2}, and not just the ones in binary systems, so that previously undetected binary systems may be discovered.

After applying these cuts, we find seven objects. Their associated millisecond pulsars are J0437-4715, J1012+5307, J1024-0719, J1732-5049, J1910+1256, J1843-1113, and J1949+3106. No binary companion has previously been detected for J1843-1113. 

We calculate the false alarm probability (FAP) for each \gaia object with a series of at least $10^6$ random trials to test the null hypothesis that the candidate binary companion is a chance alignment with the pulsar. For each trial, we inject random offsets in the pulsar's RA and DEC drawn from Gaussian distributions with a standard deviation of 3 degrees. We query \gaia for sources within $3''$ of this randomized position and make the above-described cuts on the results. The fraction of trials in which one or more sources remain after the cuts gives the FAP. We chose to randomize the pulsar's position within a subsection of the sky to reflect the density of objects surrounding the pulsar: \cm{objects in the crowded galactic plane, such as J1024-0719, are likely to have a higher FAP than those at high galactic latitude.} 

While \cite{JenningsEtAl:2018} claim a FAP between $10^{-3}$ and $10^{-5}$ for their cross-match results, they computed a FAP by looking at the number of \gaia sources with magnitudes brighter than the \gaia candidate within $1'$ of the pulsar. Our FAP calculation method, described above, enables us to make a more confident detection of binary companions.

\section{Results}

\begin{figure}
\centering     
\includegraphics[clip, trim=1.8cm 0cm 0cm 0cm, width=0.49\textwidth]{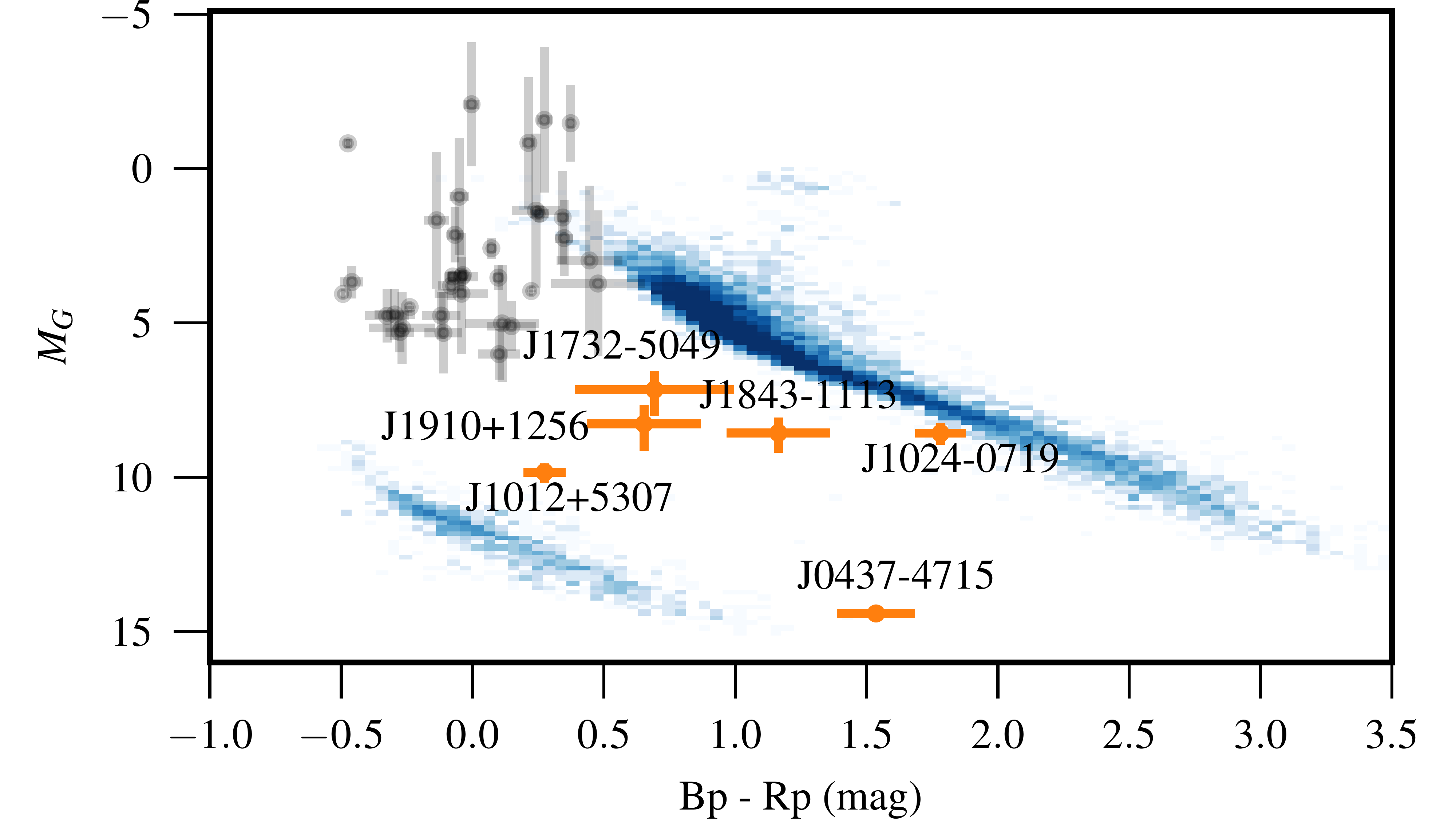}
\caption{Color-magnitude diagram of a subsample of well-measured \gaia stars (blue), where the white dwarf sequence and main sequence are visible in the lower left and upper right, respectively. Candidate binary companions to \citetalias{iptadr2} pulsars are orange. \cm{Grey points are a sample from \cite{Brown2020}'s Extremely Low Mass (ELM) survey of double white dwarf binaries, which seem to be occupy a different part of the color-magnitude diagram. }Two of the companions are consistent with the white dwarf sequence, and a few are in the valley between the white dwarf sequence and main sequence -- likely low-mass, helium-rich degenerate dwarfs \citep{ndm+04}.
}
\label{fig:HRD}
\end{figure}

We summarize our results in Table \ref{tab:results}. The seven pulsar companions are shown on the color-magnitude diagram in Figure \ref{fig:HRD}, in orange. For comparison, we also show a random subsample of well-measured white dwarf and main sequence stars as described in \citet{gaiadr2}. We dust correct the \gaia magnitudes for the pulsar companions assuming the median dust values are correct, either from the 3D dust map \cite{3Ddust}, where available, or \cite{sfd} where the 3D dust map is not available. That is to say, the uncertainties on the dust are not included in the error bars in the figure. \cm{We also plot a selection of double white dwarfs from \cite{Brown2020}, and find that they appear to be from a distinct population to the binary companions we found here.} 
Furthermore, \cite{kkn+16} and \cite{bjs+16} claim that the binary companion to J1024-0719 is a low-mass main sequence star, and this is consistent with our dust-corrected findings in Figure \ref{fig:HRD}. 

In the following subsections, we compute distances to the IPTA MSPs from \gaia and PTA parallax measurements, using the \cite{bjr+18} distance prior which includes the global parallax offset of $-0.029$~mas determined from \gaia's observations of quasars \citep{lhb+18}. While PTA measurements typically use the \cite{lk73} correction, here we use \cite{bjr+18}-corrected distances, while also providing the improved parallax measurements so that the reader may choose any distance prior they wish. \cite{lk73} and \cite{bjr+18}-corrected distances are however comparable, and a thorough comparison of these is explored in \cite{lbs18}.

\subsection{J0437-4715: an even more precise distance measurement}
Pulsar J0437-4715 has a white dwarf companion and has binary orbital period of 5.7 days~\citep{vbs+08}. Using VLBI, \cite{dvt+08} measured a parallax of $6.396 
\pm 0.054$~mas to this system, and more recently \cite{rhc+16} measured a parallax of $6.37 \pm 0.09 $~mas. Through a complementary method based on measuring the binary orbital period derivative \citep{s70}, \cite{rhc+16} also made an independent distance measurement to this binary system, and found it be $156.79\pm 0.25$~pc. 

With a strong detection of the parallax with S/N $\sim 12$, we identify J0437-4715's white dwarf companion as \gaia designation Gaia DR2 4789864076732331648 with a parallax measurement of $8.33
\pm 0.68$~mas. It is unclear as to why this parallax measurement is so poor compared to PTA measurements, since this binary system is close to the Earth.
Using the method described above, we find that the FAP of this candidate is $<7\times10^{-8}$, corresponding to a $>5\sigma$ detection.

We normalize and multiply together the independent parallax measurements from \gaia,  \cite{dvt+08}, and \cite{rhc+16}, and find only a marginal improvement in the parallax measurement: $6.40\pm 0.05$~mas, Figure \ref{fig:0437}. We then compute the distance from the parallax, and find this to be $ 155.57 \pm 1.21$~pc. Since the dynamical distance estimate is another independent way to measure the distance, we further normalize and multiply together the distance posteriors from the combined parallax measurement and \cite{rhc+16}'s dynamical distance measurement, and find a final combined distance of $156.74 \pm 0.24$~pc (the pink stripe in Figure \ref{fig:0437}). It is clear that here the \gaia data play a minimal role and only affect the distance estimate at the centiparsec level.

\citetalias{ymw+17} model's mean distance is exactly correct for this pulsar since the authors used J0437-4715 as a calibration point. We therefore cannot use this fact to infer the overall reliability of the \citetalias{ymw+17} model. 
Moreover, since the distance to this binary system is so well-known via pulsar timing and VLBI, it may also be useful as a calibration point for future \gaia data releases.

\subsection{J1012+5307: tests of General Relativity}
J1012+5307~\citep{n95} is in a $0.6$-day period orbit with a white dwarf (\citetalias{iptadr2}). Not only is this an IPTA pulsar, but it has also been used to carry out tests of General Relativity \citep{lwj+09}, looking to constrain dipolar radiation, which intrinsically relies on the binary's distance. Here we identify the white dwarf companion to J1012+5307 as Gaia DR2 851610861391010944, detected with a S/N$\sim 3$. We find the FAP of this candidate to be $<7\times10^{-8}$ -- another $>5\sigma$ detection.
\begin{figure*}[ht!]
\centering 
\subfigure[J1012+5307 Distance]
        {
		\includegraphics[scale=1]{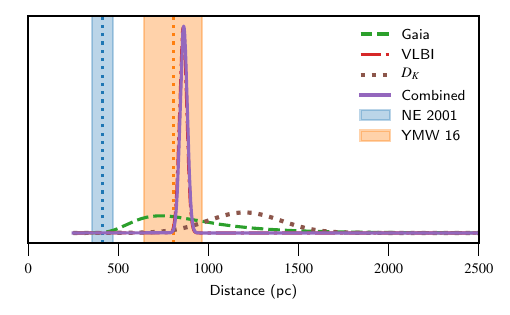} 
        }

\subfigure[J1024-0719 Distance]
        {
		\includegraphics[scale=1]{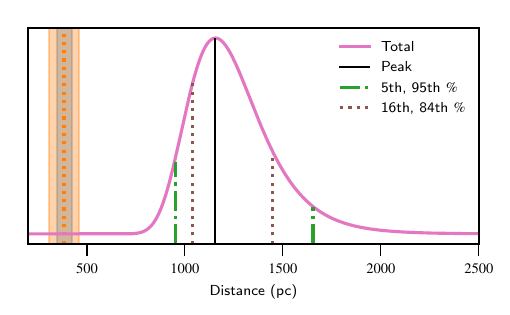}
        }
\subfigure[J1732-5049 Distance]
        {
		\includegraphics[scale=1]{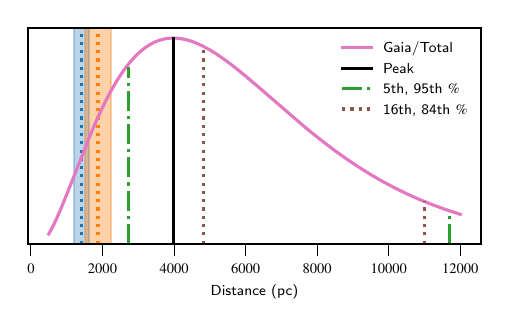}
        }
     
\subfigure[J1843-1113 Distance]
        {
		\includegraphics[scale=1]{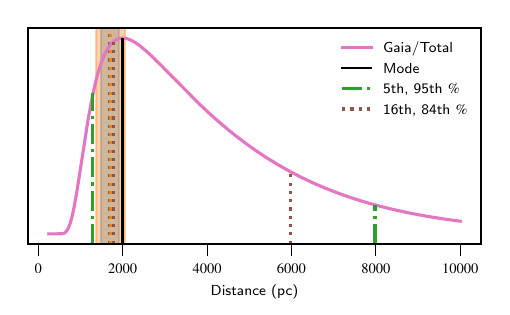} 
        }
\subfigure[J1910+1256 Distance]
        {
		\includegraphics[scale=1]{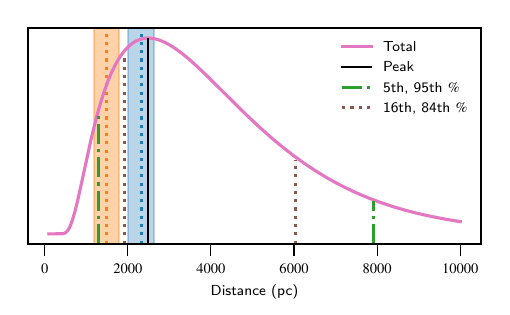} 
        }

\caption{Distance estimates to the following five binary pulsar systems. As before, DM-based distance estimates are from \citetalias{ymw+17} (yellow strip) and \citetalias{cl02} (blue strip), with their mean values represented as a dashed line. \cm{In panel (a), the distance to binary pulsar J1012+5307 is only marginally improved since this system now has a VLBI measurement \citep{Ding2020}. J1024-0719 is a very interesting ultra-wide period binary ($P>200$~yr) which we will continue to monitor. For J1732-5049 we report the first parallax measurement to this binary system. We also find a potential binary candidate for pulsar J1843-1113, which we will continue to monitor, and the low-mass white dwarf companion to J1910+1256. We identify optical companions in \gaia data to J1732-5049,  J1910+1256,  J1843-1113, and J1949+3106 for the first time.} }
\label{fig:dist}
\end{figure*}

\cm{We combine the \gaia data with the \cite{Ding2020} VLBI measurements, and find a combined parallax measurement of $1.17 \pm 0.02$~mas, which is unchanged. We add the combined parallax measurements to the recent distance estimate based on the orbital period derivative of the binary, $\dot{P}_b$ from NANOGrav~\citep{abb18}, and find the final distance to this system is $862.6 \pm 19.9$~pc.}

\subsection{J1024-0719: ultra-wide period binary}
Pulsar J1024-0719 may be in an ultra-wide period orbit with a low-mass companion. Estimated orbital periods range by orders of magnitude, from $P_b>200$~yrs \citep{bjs+16} to $2\lesssim P_b \lesssim 20$ kyr \citep{kkn+16}. Here we identify \gaia object Gaia DR2 3775277872387310208 as its binary companion. While this is a low S/N$\sim 1.25$ parallax detection, we find the FAP of this candidate to be $<7\times10^{-8}$, a $>5\sigma$ detection, and are therefore confident in this association. This object is a low-mass main sequence star, identified by \cite{kkn+16} as 1024-Br, and as 2MASS J10243869-0719190 in \cite{bjs+16}. 

By combining the \gaia parallax measurement with the PTA one we find a new and improved parallax measurement of $ 0.82\pm 0.14$~mas. The distance to this binary is  $1155^{+294}_{-116}$~pc. Since the distance posterior is slightly skewed, we report the peak of the distribution as the point value, and errors are 84th and 16th percentile. This is $ 1155^{+499}_{-205}$~pc if we use the 95th and 5th percentiles. 

\subsection{J1732-5049: first parallax measurement}
We identify the companion to J1732-5049 \citep{rb01} as Gaia DR2 5946288492263176704, with a FAP of $5.51\times 10^{-3}$~(a $2.8\sigma$ detection). While the RA and DEC of the binary companion are reasonable (offset from IPTA data by $1.6\sigma$ in RA and $0.2\sigma$ in DEC), the parallax measurement of $-1.18\pm2.84$~mas is negative, and thus this distance estimate is heavily influenced by the \citealt{bjr+18} distance prior. Future data releases should improve the parallax and proper motion measurements to this object, which will improve our confidence in this association, or disfavor it. 

The posterior distribution for the distance is very broad. Since the 16th percentile value is larger than the peak of the distribution due to the fast rise of the posterior, we write our best distance estimate as $3980~(4820, 10999)$~pc: here the point value is the peak of the distribution, the distance at the 84th percentile is +10999~pc and at the 16th percentile is 4820~pc. This is $3980^{+7716}_{-1253}$~pc for the the 95th and 5th percentiles.
Since the distance estimate here is still quite poor, we will defer discussing the implications of this binary's detection to future work.

\subsection{J1843-1113: new binary candidate}
Pulsar J1843-1113 has no previously identified binary companion. The \gaia cross-match returns an object with a parallax S/N$\sim 1.2$ that passes all of our consistency checks, Gaia ID 4106823440438736384, with a FAP of $1.19\times 10^{-2}$ ($2.5\sigma$ {\it detection}). Using \gaia's parallax measurement to this new-found companion of $0.79\pm 0.64$~mas, and combining this with the PTA parallax measurement of the pulsar, we determine its distance is  $1997^{+3972}_{-205}$~pc. The 95th and 5th percentiles are $1997^{+5983}_{-712}$~pc.

We will monitor this binary candidate in future data releases to see how the parallax improves, and defer further discussion until then.

\subsection{J1910+1256: a tentative binary candidate}
Pulsar J1910+1256 \citep{sfl+05} is a binary MSP with a low-mass white dwarf companion with a minimum mass of $0.18~M_\odot$ \citep{dcl+16}. We tentatively identify its white dwarf as Gaia~DR2~4314046781982561920. However, this object has a negative parallax measurement of $-0.42\pm0.80$~mas, hence $|\mathrm{S/N}|\sim0.5$. While its FAP is $1.61\times 10^{-2}$ (a weak $2.4\sigma$ {\it detection}), as \gaia proper motion measurements improve we will be able to either make a more confident detection of this object, or rule it out completely.
Combining the PTA and \gaia parallaxes yields a new parallax measurement of $0.58\pm 0.54$~mas, or a distance of $2491^{+3559}_{-568}$~pc. The distance is $2491^{+5417}_{-1191}$~pc if considering the 95th and 5th percentiles. 
\begin{table*}[t]
\begin{center}
\begin{tabular}
{lllllllr}
\hline\hline
Pulsar 		&  D$_{DM}$ (pc) & D$_{DM}$ (pc)  & Pulsar        & \gaia           & Combined          &   Combined & Reference \\
			& NE$2001$ 	     &YMW$16^\dagger$ &parallax (mas) & parallax (mas)  & parallax (mas)   & Distance (pc)& \\
\hline
 J0437-4715 & $139^{+33}_{-29}$ & 156.1 & $6.37 \pm 0.09$ 	& $8.35 \pm 0.69$	&	$6.40\pm 0.05$ 	&$156.74 \pm 0.24$  & \citetalias{dcl+16}, \citetalias{dvt+08}\\
 J1012+5307 & $411^{+59}_{-56}$	& 804.5 & $1.17 \pm 0.02$	    & $1.32 \pm 0.41$	&	$1.17 \pm 0.02$ 	    &$862.6 \pm 19.9$ & \cite{Ding2020}\\
 J1024-0719 & $386^{+39}_{-38}$	& 382.3	& $0.86 \pm 0.15$ & $ 0.53 \pm 0.42 $	&	$ 0.82 \pm 0.14$ 	& $ 1155^{+294}_{-116}$	&	\citetalias{iptadr2}, \citetalias{kkn+16}, \citetalias{dcl+16}\\
J1732-5049 &	$1411^{+206}_{-198}$	&	$1875.0$ 	& None	& $-1.18\pm2.84$	& $-1.18\pm2.84$	& $3980~(4820, 10999)$	&	\citetalias{iptadr2}\\
 J1843-1113 &	$1700\pm 300$	&	$1707.0$ 	& $0.62\pm  0.34$	& $0.79 \pm 0.64$	& $0.66\pm 0.33$ 	& $ 1997^{+3972}_{-205}$ &  	\citetalias{iptadr2}\\
J1910+1256 	&  $2327^{+311}_{-317}$	&1496.0 		& $1.44 \pm 0.74$	& $-0.42 \pm 0.80$ & $0.58 \pm 0.54$	& $2491^{+3559}_{-568}$ &	\citetalias{iptadr2, dcl+16}\\
 J1949+3106 &	$6514^{+822}_{-965}$	&	$7466.3$ 	& $-0.67 \pm 3.18$	& $0.98 \pm 0.35$	& $0.96 \pm 0.35$	& $1295~(1306, 5399)$	& \citetalias{iptadr2}\\
 \hline \hline
\end{tabular}
\end{center}
\caption{Summary of results. Here we report distance measurements to binary pulsars in terms of their dispersion measure estimate, the pulsar timing parallax measurement, the \gaia parallax measurement to their companion, their combined parallax and subsequent inferred distance. We compute the final distance by combining the parallax-based distance estimate with dynamical distance estimates, when available, since these are also independent distance estimates. Dispersion measures were reported in \citetalias{iptadr2} and references therein, and pulsar distances were calculated based on the DM via the online tools for \citetalias{cl02} and \citetalias{ymw+17}. DM-based distances are shown for comparison, and are not used to calculate combined distances. $^\dagger$As in the literature, we assume the error in the distance from the model is $\pm 20\%$. Asymmetric error bars in the distance are the 16th and 84th percentiles. Values in brackets are also 16th and 84th percentiles, but the 16th percentile is greater than the peak, so this cannot be written in the usual way.}
\label{tab:results}
\end{table*}

\subsection{J1949+3106: best parallax measurement}
The dispersion measure of pulsar J1949+3106 \citep{dfc+12} leads us to believe that it is a distant binary system. In fact, it has one of the highest dispersion measures of any millisecond pulsar, a companion with a mass of $\sim 1 M_\odot$\citep{dfc+12}, and has an uncertain parallax measurement of $-0.6(3.2)$ mas. Here we identify its binary companion in \gaia as Gaia DR2 2033684263247409920. We find this \gaia object has a parallax measurement of $0.97 \pm 0.35$, and is detected with  $S/N = 2.8$, making this the best known parallax measurement to this binary system. We also find that the FAP is $3.19\times 10^{-3}$, a $3\sigma$ detection.

We combine the parallax posteriors and find an updated parallax measurement of $0.96 \pm 0.35 $~mas, and compute the distance. Since the posterior distribution for the distance is highly skewed, we report the the peak at 1295~pc as a point estimate, while the 16th percentile value is 1306~pc and the 84th percentile is 5399~pc. We write this in Table 1 as $1295~(1306, 5399)$~pc, and report that the 95th and 5th percentile measurements are $1295^{+6413}_{-330}$~pc. 

This is a very intriguing binary. DM-based estimates put this system $6514^{+822}_{-965}$~pc (NE2001) and $ 7466 \pm 1493$~pc (YMW16) -- a factor of $\sim 5$ further than expected from our parallax-based measurement. This may be pointing to a fault in both these independent electron density models of the galaxy, or that there is still a cloud of material surrounding this binary system. Indeed, this DM-distance overestimate combined with J1949+3106's relatively poor timing precision \citep{abb18} may suggest that there is additional material surrounding the binary, possibly from a recent mass-transfer phase, which is creating a local DM contribution in addition to the galactic DM.

\section{Discussion}
We find that the errors on the DM distance estimates are typically underestimated. In the case of J1024-0719 and 1732-5049, the binary systems are more than factor of two further than both NE2001 and YMW16 predict, pointing to an underestimate in these electron density estimates in the direction of the binaries. Conversely, J1949+3106 is a factor of 5 closer than both DM models predict, which may point to an additional, local DM contribution from around the binary system, possibly from a recent mass-transfer phase. \cm{DM errors are also not straightforward, and deviations from the frequently quoted $\sim 20\%$ value is not entirely unexpected since DM models are more accurate at low latitudes for young, distant pulsars, while at high latitude there are fewer calibration points \citep{Cordes2016}}. 

In Figure \ref{fig:HRD}, we see that there are a few companions with a brightness that is consistent with a young white dwarf, but a color that places it in the valley between the white dwarf sequence and main sequence. Though these stars are only $2$ to $3\sigma$ from the main sequence, a non-trivial population of stars lie in this region of the color-magnitude diagram from the full \gaia catalog \citep{gaiahrd}. According to isochrones in \cite{ndm+04}, these are likely helium-rich degenerate dwarfs. \cm{Our companions also occupy a distinct part of the color-magnitude diagram than the double white dwarf sample from \cite{Brown2020} ELM survey, which is also interesting.}

\gaia's parallax measurements improve as $T^{-1/2}$, therefore this will be decrease current parallax errors by a factor of $(38/12)^{-1/2}\sim 0.6$ -- also noted by \cite{JenningsEtAl:2018} -- in \gaia's early Data Release 3 (EDR3; \citealt{GaiaEDR3}). A companion paper (Moran et al., in prep) will examine a new cross-match between IPTA data and \gaia EDR3, since the full data release will not further improve parallax measurements. 

While PTA timing parallax errors are still large, kinematic distance measurements from the binary orbital period derivative scale as $\dot{P}_b \propto T^{-5/2}$~\citep{bb96}. Therefore over 38 months, for systems with $\dot{P}_b$ measurements, errors in $D_K$ should be $(38/12)^{-5/2}\sim 0.06$ of their present value. This is especially encouraging for binary systems such as J1012+5307, where the distance constraint now makes use of both these measurements. While it is clear that the kinematic distance measurement will eventually be the tighter distance constraint, few MSPs have $\dot{P}_b$ measurements (this is a difficult measurement). Moreover, having multiple independent distance measurements will certainly instill greater confidence in the results, which is important for making claims of the detection of dipolar radiation \citep{lwj+09}, claiming changes in Newton's constant $G$~\citep{lyp+18}, and the eventual detection of continuous nanohertz gravitational waves within the next decade~\citep{mls+17, Xin2021}.

\section*{Acknowledgments}
We are grateful for the free Python software provided  by \cite{scipy, numpy, astropy}. The authors thank W. M. Farr, D. W. Hogg, M. Bailes, Y. Levin, D. Kaplan, D. Nice, R. Smart, A. Price-Whelan, M. Pitkin, L. A. Nelson, and J. Hernandez for useful conversations. We thank R. N. Manchester for clarifications on the YMW16 model. This work has made use of data from the European Space Agency (ESA) mission
{\it Gaia} (\url{https://www.cosmos.esa.int/gaia}), processed by the {\it Gaia}
Data Processing and Analysis Consortium (DPAC,
\url{https://www.cosmos.esa.int/web/gaia/dpac/consortium}). Funding for the DPAC
has been provided by national institutions, in particular the institutions
participating in the \gaia Multilateral Agreement. 
This project was carried out in part at the 2018 \gaia Sprint, hosted by the Center for Computational Astrophysics of the Flatiron Institute in New York City. The Flatiron Institute is supported by the Simons Foundation. This research was supported in part by the National Science Foundation under Grants No. NSF PHY-1748958, PHY-2020265, and AST-2106552.


\begin{thebibliography}{}
\expandafter\ifx\csname natexlab\endcsname\relax\def\natexlab#1{#1}\fi
\providecommand{\url}[1]{\href{#1}{#1}}

\bibitem[{{Aggarwal} {et~al.}(2019){Aggarwal}, {Arzoumanian}, {Baker},
  {Brazier}, {Brinson}, {Brook}, {Burke-Spolaor}, {Chatterjee}, {Cordes},
  {Cornish}, {Crawford}, {Crowter}, {Cromartie}, {DeCesar}, {Demorest},
  {Dolch}, {Ellis}, {Ferdman}, {Ferrara}, {Fonseca}, {Garver-Daniels},
  {Gentile}, {Hazboun}, {Holgado}, {Huerta}, {Islo}, {Jennings}, {Jones},
  {Jones}, {Kaiser}, {Kaplan}, {Kelley}, {Key}, {Lam}, {Lazio}, {Levin},
  {Lorimer}, {Luo}, {Lynch}, {Madison}, {McLaughlin}, {McWilliams},
  {Mingarelli}, {Ng}, {Nice}, {Pennucci}, {Pol}, {Ransom}, {Ray}, {Siemens},
  {Simon}, {Spiewak}, {Stairs}, {Stinebring}, {Stovall}, {Swiggum}, {Taylor},
  {Turner}, {Vallisneri}, {van Haasteren}, {Vigeland }, {Witt}, {Zhu}, \& {(The
  NANOGrav Collaboration}}]{cgw19}
{Aggarwal}, K., {Arzoumanian}, Z., {Baker}, P.~T., {et~al.} 2019, \apj, 880,
  116

\bibitem[{{Arzoumanian} {et~al.}(2018){Arzoumanian}, {Brazier},
  {Burke-Spolaor}, {Chamberlin}, {Chatterjee}, {Christy}, {Cordes}, {Cornish},
  {Crawford}, {Thankful Cromartie}, {Crowter}, {DeCesar}, {Demorest}, {Dolch},
  {Ellis}, {Ferdman}, {Ferrara}, {Fonseca}, {Garver-Daniels}, {Gentile},
  {Halmrast}, {Huerta}, {Jenet}, {Jessup}, {Jones}, {Jones}, {Kaplan}, {Lam},
  {Lazio}, {Levin}, {Lommen}, {Lorimer}, {Luo}, {Lynch}, {Madison}, {Matthews},
  {McLaughlin}, {McWilliams}, {Mingarelli}, {Ng}, {Nice}, {Pennucci}, {Ransom},
  {Ray}, {Siemens}, {Simon}, {Spiewak}, {Stairs}, {Stinebring}, {Stovall},
  {Swiggum}, {Taylor}, {Vallisneri}, {van Haasteren}, {Vigeland}, {Zhu}, \&
  {NANOGrav Collaboration}}]{abb18}
{Arzoumanian}, Z., {Brazier}, A., {Burke-Spolaor}, S., {et~al.} 2018, \apjs,
  235, 37

\bibitem[{{Arzoumanian} {et~al.}(2020){Arzoumanian}, {Baker}, {Blumer},
  {B{\'e}csy}, {Brazier}, {Brook}, {Burke-Spolaor}, {Chatterjee}, {Chen},
  {Cordes}, {Cornish}, {Crawford}, {Cromartie}, {Decesar}, {Demorest}, {Dolch},
  {Ellis}, {Ferrara}, {Fiore}, {Fonseca}, {Garver-Daniels}, {Gentile}, {Good},
  {Hazboun}, {Holgado}, {Islo}, {Jennings}, {Jones}, {Kaiser}, {Kaplan},
  {Kelley}, {Key}, {Laal}, {Lam}, {Lazio}, {Lorimer}, {Luo}, {Lynch},
  {Madison}, {McLaughlin}, {Mingarelli}, {Ng}, {Nice}, {Pennucci}, {Pol},
  {Ransom}, {Ray}, {Shapiro-Albert}, {Siemens}, {Simon}, {Spiewak}, {Stairs},
  {Stinebring}, {Stovall}, {Sun}, {Swiggum}, {Taylor}, {Turner}, {Vallisneri},
  {Vigeland}, {Witt}, \& {Nanograv Collaboration}}]{ng12p5}
{Arzoumanian}, Z., {Baker}, P.~T., {Blumer}, H., {et~al.} 2020, \apjl, 905, L34

\bibitem[{{Babak} {et~al.}(2016){Babak}, {Petiteau}, {Sesana}, {Brem},
  {Rosado}, {Taylor}, {Lassus}, {Hessels}, {Bassa}, {Burgay}, {Caballero},
  {Champion}, {Cognard}, {Desvignes}, {Gair}, {Guillemot}, {Janssen},
  {Karuppusamy}, {Kramer}, {Lazarus}, {Lee}, {Lentati}, {Liu}, {Mingarelli},
  {Os{\l}owski}, {Perrodin}, {Possenti}, {Purver}, {Sanidas}, {Smits},
  {Stappers}, {Theureau}, {Tiburzi}, {van Haasteren}, {Vecchio}, \&
  {Verbiest}}]{babak+16}
{Babak}, S., {Petiteau}, A., {Sesana}, A., {et~al.} 2016, \mnras, 455, 1665

\bibitem[{{Backer} {et~al.}(1982){Backer}, {Kulkarni}, {Heiles}, {Davis}, \&
  {Goss}}]{bkh+82}
{Backer}, D.~C., {Kulkarni}, S.~R., {Heiles}, C., {Davis}, M.~M., \& {Goss},
  W.~M. 1982, \nat, 300, 615

\bibitem[{{Bailer-Jones} {et~al.}(2018){Bailer-Jones}, {Rybizki}, {Fouesneau},
  {Mantelet}, \& {Andrae}}]{bjr+18}
{Bailer-Jones}, C.~A.~L., {Rybizki}, J., {Fouesneau}, M., {Mantelet}, G., \&
  {Andrae}, R. 2018, \aj, 156, 58

\bibitem[{{Bassa} {et~al.}(2016){Bassa}, {Janssen}, {Stappers}, {Tauris},
  {Wevers}, {Jonker}, {Lentati}, {Verbiest}, {Desvignes}, {Graikou},
  {Guillemot}, {Freire}, {Lazarus}, {Caballero}, {Champion}, {Cognard},
  {Jessner}, {Jordan}, {Karuppusamy}, {Kramer}, {Lazaridis}, {Lee}, {Liu},
  {Lyne}, {McKee}, {Os{\l}owski}, {Perrodin}, {Sanidas}, {Shaifullah}, {Smits},
  {Theureau}, {Tiburzi}, \& {Zhu}}]{bjs+16}
{Bassa}, C.~G., {Janssen}, G.~H., {Stappers}, B.~W., {et~al.} 2016, \mnras,
  460, 2207

\bibitem[{{Bell} \& {Bailes}(1996)}]{bb96}
{Bell}, J.~F., \& {Bailes}, M. 1996, \apj, 456, L33

\bibitem[{{Bhattacharya} \& {van den Heuvel}(1991)}]{bvdh91}
{Bhattacharya}, D., \& {van den Heuvel}, E.~P.~J. 1991, \physrep, 203, 1

\bibitem[{{Brown} {et~al.}(2020){Brown}, {Kilic}, {Kosakowski}, {Andrews},
  {Heinke}, {Ag{\"u}eros}, {Camilo}, {Gianninas}, {Hermes}, \&
  {Kenyon}}]{Brown2020}
{Brown}, W.~R., {Kilic}, M., {Kosakowski}, A., {et~al.} 2020, \apj, 889, 49

\bibitem[{{Cordes} \& {Lazio}(2002)}]{cl02}
{Cordes}, J.~M., \& {Lazio}, T.~J.~W. 2002, ArXiv Astrophysics e-prints,
  astro-ph/0207156

\bibitem[{{Cordes} {et~al.}(2016){Cordes}, {Shannon}, \&
  {Stinebring}}]{Cordes2016}
{Cordes}, J.~M., {Shannon}, R.~M., \& {Stinebring}, D.~R. 2016, \apj, 817, 16

\bibitem[{{Deller} {et~al.}(2008){Deller}, {Verbiest}, {Tingay}, \&
  {Bailes}}]{dvt+08}
{Deller}, A.~T., {Verbiest}, J.~P.~W., {Tingay}, S.~J., \& {Bailes}, M. 2008,
  ApJL, 685, L67

\bibitem[{{Deneva} {et~al.}(2012){Deneva}, {Freire}, {Cordes}, {Lyne},
  {Ransom}, {Cognard}, {Camilo}, {Nice}, {Stairs}, {Allen}, {Bhat}, {Bogdanov},
  {Brazier}, {Champion}, {Chatterjee}, {Crawford}, {Desvignes}, {Hessels},
  {Jenet}, {Kaspi}, {Knispel}, {Kramer}, {Lazarus}, {van Leeuwen}, {Lorimer},
  {Lynch}, {McLaughlin}, {Scholz}, {Siemens}, {Stappers}, {Stovall}, \&
  {Venkataraman}}]{dfc+12}
{Deneva}, J.~S., {Freire}, P.~C.~C., {Cordes}, J.~M., {et~al.} 2012, \apj, 757,
  89

\bibitem[{{Desvignes} {et~al.}(2016){Desvignes}, {Caballero}, {Lentati},
  {Verbiest}, {Champion}, {Stappers}, {Janssen}, {Lazarus}, {Os{\l}owski},
  {Babak}, {Bassa}, {Brem}, {Burgay}, {Cognard}, {Gair}, {Graikou},
  {Guillemot}, {Hessels}, {Jessner}, {Jordan}, {Karuppusamy}, {Kramer},
  {Lassus}, {Lazaridis}, {Lee}, {Liu}, {Lyne}, {McKee}, {Mingarelli},
  {Perrodin}, {Petiteau}, {Possenti}, {Purver}, {Rosado}, {Sanidas}, {Sesana},
  {Shaifullah}, {Smits}, {Taylor}, {Theureau}, {Tiburzi}, {van Haasteren}, \&
  {Vecchio}}]{dcl+16}
{Desvignes}, G., {Caballero}, R.~N., {Lentati}, L., {et~al.} 2016, \mnras, 458,
  3341

\bibitem[{{Detweiler}(1979)}]{det79}
{Detweiler}, S. 1979, ApJ, 234, 1100

\bibitem[{{Ding} {et~al.}(2020){Ding}, {Deller}, {Freire}, {Kaplan}, {Lazio},
  {Shannon}, \& {Stappers}}]{Ding2020}
{Ding}, H., {Deller}, A.~T., {Freire}, P., {et~al.} 2020, \apj, 896, 85

\bibitem[{{Edwards} \& {Bailes}(2001)}]{rb01}
{Edwards}, R.~T., \& {Bailes}, M. 2001, \apj, 547, L37

\bibitem[{{Gaia Collaboration} {et~al.}(2018{\natexlab{a}}){Gaia
  Collaboration}, {Brown}, {Vallenari}, {Prusti}, {de Bruijne}, {Babusiaux},
  {Bailer-Jones}, {Biermann}, {Evans}, {Eyer}, \& et~al.}]{gaiadr2}
{Gaia Collaboration}, {Brown}, A.~G.~A., {Vallenari}, A., {et~al.}
  2018{\natexlab{a}}, \aap, 616, A1

\bibitem[{{Gaia Collaboration} {et~al.}(2018{\natexlab{b}}){Gaia
  Collaboration}, {Babusiaux}, {van Leeuwen}, {Barstow}, {Jordi}, {Vallenari},
  {Bossini}, {Bressan}, {Cantat-Gaudin}, {van Leeuwen}, {Brown}, {Prusti}, {de
  Bruijne}, {Bailer-Jones}, {Biermann}, {Evans}, {Eyer}, {Jansen}, {Klioner},
  {Lammers}, {Lindegren}, {Luri}, {Mignard}, {Panem}, {Pourbaix}, {Randich},
  {Sartoretti}, {Siddiqui}, {Soubiran}, {Walton}, {Arenou}, {Bastian},
  {Cropper}, {Drimmel}, {Katz}, {Lattanzi}, {Bakker}, {Cacciari},
  {Casta{\~n}eda}, {Chaoul}, {Cheek}, {De Angeli}, {Fabricius}, {Guerra},
  {Holl}, {Masana}, {Messineo}, {Mowlavi}, {Nienartowicz}, {Panuzzo},
  {Portell}, {Riello}, {Seabroke}, {Tanga}, {Th{\'e}venin}, {Gracia-Abril},
  {Comoretto}, {Garcia-Reinaldos}, {Teyssier}, {Altmann}, {Andrae}, {Audard},
  {Bellas-Velidis}, {Benson}, {Berthier}, {Blomme}, {Burgess}, {Busso},
  {Carry}, {Cellino}, {Clementini}, {Clotet}, {Creevey}, {Davidson}, {De
  Ridder}, {Delchambre}, {Dell'Oro}, {Ducourant}, {Fern{\'a}ndez-
  Hern{\'a}ndez}, {Fouesneau}, {Fr{\'e}mat}, {Galluccio}, {Garc{\'\i}a-Torres},
  {Gonz{\'a}lez-N{\'u}{\~n}ez}, {Gonz{\'a}lez-Vidal}, {Gosset}, {Guy},
  {Halbwachs}, {Hambly}, {Harrison}, {Hern{\'a}ndez}, {Hestroffer}, {Hodgkin},
  {Hutton}, {Jasniewicz}, {Jean-Antoine-Piccolo}, {Jordan}, {Korn},
  {Krone-Martins}, {Lanzafame}, {Lebzelter}, {L{\"o}ffler}, {Manteiga},
  {Marrese}, {Mart{\'\i}n-Fleitas}, {Moitinho}, {Mora}, {Muinonen}, {Osinde},
  {Pancino}, {Pauwels}, {Petit}, {Recio-Blanco}, {Richards}, {Rimoldini},
  {Robin}, {Sarro}, {Siopis}, {Smith}, {Sozzetti}, {S{\"u}veges}, {Torra}, {van
  Reeven}, {Abbas}, {Abreu Aramburu}, {Accart}, {Aerts}, {Altavilla},
  {{\'A}lvarez}, {Alvarez}, {Alves}, {Anderson}, {Andrei}, {Anglada Varela},
  {Antiche}, {Antoja}, {Arcay}, {Astraatmadja}, {Bach}, {Baker},
  {Balaguer-N{\'u}{\~n}ez}, {Balm}, {Barache}, {Barata}, {Barbato}, {Barblan},
  {Barklem}, {Barrado}, {Barros}, {Bartholom{\'e} Mu{\~n}oz}, {Bassilana},
  {Becciani}, {Bellazzini}, {Berihuete}, {Bertone}, {Bianchi}, {Bienaym{\'e}},
  {Blanco-Cuaresma}, {Boch}, {Boeche}, {Bombrun}, {Borrachero}, {Bouquillon},
  {Bourda}, {Bragaglia}, {Bramante}, {Breddels}, {Brouillet},
  {Br{\"u}semeister}, {Brugaletta}, {Bucciarelli}, {Burlacu}, {Busonero},
  {Butkevich}, {Buzzi}, {Caffau}, {Cancelliere}, {Cannizzaro}, {Carballo},
  {Carlucci}, {Carrasco}, {Casamiquela}, {Castellani}, {Castro-Ginard},
  {Charlot}, {Chemin}, {Chiavassa}, {Cocozza}, {Costigan}, {Cowell}, {Crifo},
  {Crosta}, {Crowley}, {Cuypers}, {Dafonte}, {Damerdji}, {Dapergolas}, {David},
  {David}, {de Laverny}, {De Luise}, {De March}, {de Martino}, {de Souza}, {de
  Torres}, {Debosscher}, {del Pozo}, {Delbo}, {Delgado}, {Delgado}, {Diakite},
  {Diener}, {Distefano}, {Dolding}, {Drazinos}, {Dur{\'a}n}, {Edvardsson},
  {Enke}, {Eriksson}, {Esquej}, {Eynard Bontemps}, {Fabre}, {Fabrizio},
  {Faigler}, {Falc{\~a}o}, {Farr{\`a}s Casas}, {Federici}, {Fedorets},
  {Fernique}, {Figueras}, {Filippi}, {Findeisen}, {Fonti}, {Fraile}, {Fraser},
  {Fr{\'e}zouls}, {Gai}, {Galleti}, {Garabato}, {Garc{\'\i}a-Sedano},
  {Garofalo}, {Garralda}, {Gavel}, {Gavras}, {Gerssen}, {Geyer}, {Giacobbe},
  {Gilmore}, {Girona}, {Giuffrida}, {Glass}, {Gomes}, {Granvik}, {Gueguen},
  {Guerrier}, {Guiraud}, {Guti{\'e}}, {Haigron}, {Hatzidimitriou}, {Hauser},
  {Haywood}, {Heiter}, {Helmi}, {Heu}, {Hilger}, {Hobbs}, {Hofmann}, {Holland},
  {Huckle}, {Hypki}, {Icardi}, {Jan{\ss}en}, {Jevardat de Fombelle}, {Jonker},
  {Juh{\'a}sz}, {Julbe}, {Karampelas}, {Kewley}, {Klar}, {Kochoska}, {Kohley},
  {Kolenberg}, {Kontizas}, {Kontizas}, {Koposov}, {Kordopatis},
  {Kostrzewa-Rutkowska}, {Koubsky}, {Lambert}, {Lanza}, {Lasne}, {Lavigne}, {Le
  Fustec}, {Le Poncin-Lafitte}, {Lebreton}, {Leccia}, {Leclerc},
  {Lecoeur-Taibi}, {Lenhardt}, {Leroux}, {Liao}, {Licata}, {Lindstr{\o}m},
  {Lister}, {Livanou}, {Lobel}, {L{\'o}pez}, {Managau}, {Mann}, {Mantelet},
  {Marchal}, {Marchant}, {Marconi}, {Marinoni}, {Marschalk{\'o}}, {Marshall},
  {Martino}, {Marton}, {Mary}, {Massari}, {Matijevi{\v{c}}}, {Mazeh},
  {McMillan}, {Messina}, {Michalik}, {Millar}, {Molina}, {Molinaro},
  {Moln{\'a}r}, {Montegriffo}, {Mor}, {Morbidelli}, {Morel}, {Morris},
  {Mulone}, {Muraveva}, {Musella}, {Nelemans}, {Nicastro}, {Noval},
  {O'Mullane}, {Ord{\'e}novic}, {Ord{\'o}{\~n}ez-Blanco}, {Osborne}, {Pagani},
  {Pagano}, {Pailler}, {Palacin}, {Palaversa}, {Panahi}, {Pawlak},
  {Piersimoni}, {Pineau}, {Plachy}, {Plum}, {Poggio}, {Poujoulet},
  {Pr{\v{s}}a}, {Pulone}, {Racero}, {Ragaini}, {Rambaux}, {Ramos-Lerate},
  {Regibo}, {Reyl{\'e}}, {Riclet}, {Ripepi}, {Riva}, {Rivard}, {Rixon},
  {Roegiers}, {Roelens}, {Romero-G{\'o}mez}, {Rowell}, {Royer}, {Ruiz-Dern},
  {Sadowski}, {Sagrist{\`a} Sell{\'e}s}, {Sahlmann}, {Salgado}, {Salguero},
  {Sanna}, {Santana- Ros}, {Sarasso}, {Savietto}, {Schultheis}, {Sciacca},
  {Segol}, {Segovia}, {S{\'e}gransan}, {Shih}, {Siltala}, {Silva}, {Smart},
  {Smith}, {Solano}, {Solitro}, {Sordo}, {Soria Nieto}, {Souchay}, {Spagna},
  {Spoto}, {Stampa}, {Steele}, {Steidelm{\"u}ller}, {Stephenson}, {Stoev},
  {Suess}, {Surdej}, {Szabados}, {Szegedi-Elek}, {Tapiador}, {Taris}, {Tauran},
  {Taylor}, {Teixeira}, {Terrett}, {Teyssandier}, {Thuillot}, {Titarenko},
  {Torra Clotet}, {Turon}, {Ulla}, {Utrilla}, {Uzzi}, {Vaillant}, {Valentini},
  {Valette}, {van Elteren}, {Van Hemelryck}, {Vaschetto}, {Vecchiato},
  {Veljanoski}, {Viala}, {Vicente}, {Vogt}, {von Essen}, {Voss}, {Votruba},
  {Voutsinas}, {Walmsley}, {Weiler}, {Wertz}, {Wevers}, {Wyrzykowski},
  {Yoldas}, {{\v{Z}}erjal}, {Ziaeepour}, {Zorec}, {Zschocke}, {Zucker},
  {Zurbach}, \& {Zwitter}}]{gaiahrd}
{Gaia Collaboration}, {Babusiaux}, C., {van Leeuwen}, F., {et~al.}
  2018{\natexlab{b}}, \aap, 616, A10

\bibitem[{{Gaia Collaboration} {et~al.}(2021){Gaia Collaboration}, {Brown},
  {Vallenari}, {Prusti}, {de Bruijne}, {Babusiaux}, {Biermann}, {Creevey},
  {Evans}, {Eyer}, {Hutton}, {Jansen}, {Jordi}, {Klioner}, {Lammers},
  {Lindegren}, {Luri}, {Mignard}, {Panem}, {Pourbaix}, {Randich}, {Sartoretti},
  {Soubiran}, {Walton}, {Arenou}, {Bailer-Jones}, {Bastian}, {Cropper},
  {Drimmel}, {Katz}, {Lattanzi}, {van Leeuwen}, {Bakker}, {Cacciari},
  {Casta{\~n}eda}, {De Angeli}, {Ducourant}, {Fabricius}, {Fouesneau},
  {Fr{\'e}mat}, {Guerra}, {Guerrier}, {Guiraud}, {Jean-Antoine Piccolo},
  {Masana}, {Messineo}, {Mowlavi}, {Nicolas}, {Nienartowicz}, {Pailler},
  {Panuzzo}, {Riclet}, {Roux}, {Seabroke}, {Sordo}, {Tanga}, {Th{\'e}venin},
  {Gracia-Abril}, {Portell}, {Teyssier}, {Altmann}, {Andrae}, {Bellas-Velidis},
  {Benson}, {Berthier}, {Blomme}, {Brugaletta}, {Burgess}, {Busso}, {Carry},
  {Cellino}, {Cheek}, {Clementini}, {Damerdji}, {Davidson}, {Delchambre},
  {Dell'Oro}, {Fern{\'a}ndez-Hern{\'a}ndez}, {Galluccio}, {Garc{\'\i}a-Lario},
  {Garcia-Reinaldos}, {Gonz{\'a}lez-N{\'u}{\~n}ez}, {Gosset}, {Haigron},
  {Halbwachs}, {Hambly}, {Harrison}, {Hatzidimitriou}, {Heiter},
  {Hern{\'a}ndez}, {Hestroffer}, {Hodgkin}, {Holl}, {Jan{\ss}en}, {Jevardat de
  Fombelle}, {Jordan}, {Krone-Martins}, {Lanzafame}, {L{\"o}ffler}, {Lorca},
  {Manteiga}, {Marchal}, {Marrese}, {Moitinho}, {Mora}, {Muinonen}, {Osborne},
  {Pancino}, {Pauwels}, {Petit}, {Recio-Blanco}, {Richards}, {Riello},
  {Rimoldini}, {Robin}, {Roegiers}, {Rybizki}, {Sarro}, {Siopis}, {Smith},
  {Sozzetti}, {Ulla}, {Utrilla}, {van Leeuwen}, {van Reeven}, {Abbas}, {Abreu
  Aramburu}, {Accart}, {Aerts}, {Aguado}, {Ajaj}, {Altavilla}, {{\'A}lvarez},
  {{\'A}lvarez Cid-Fuentes}, {Alves}, {Anderson}, {Anglada Varela}, {Antoja},
  {Audard}, {Baines}, {Baker}, {Balaguer-N{\'u}{\~n}ez}, {Balbinot}, {Balog},
  {Barache}, {Barbato}, {Barros}, {Barstow}, {Bartolom{\'e}}, {Bassilana},
  {Bauchet}, {Baudesson-Stella}, {Becciani}, {Bellazzini}, {Bernet}, {Bertone},
  {Bianchi}, {Blanco-Cuaresma}, {Boch}, {Bombrun}, {Bossini}, {Bouquillon},
  {Bragaglia}, {Bramante}, {Breedt}, {Bressan}, {Brouillet}, {Bucciarelli},
  {Burlacu}, {Busonero}, {Butkevich}, {Buzzi}, {Caffau}, {Cancelliere},
  {C{\'a}novas}, {Cantat-Gaudin}, {Carballo}, {Carlucci}, {Carnerero},
  {Carrasco}, {Casamiquela}, {Castellani}, {Castro-Ginard}, {Castro Sampol},
  {Chaoul}, {Charlot}, {Chemin}, {Chiavassa}, {Cioni}, {Comoretto}, {Cooper},
  {Cornez}, {Cowell}, {Crifo}, {Crosta}, {Crowley}, {Dafonte}, {Dapergolas},
  {David}, {David}, {de Laverny}, {De Luise}, {De March}, {De Ridder}, {de
  Souza}, {de Teodoro}, {de Torres}, {del Peloso}, {del Pozo}, {Delbo},
  {Delgado}, {Delgado}, {Delisle}, {Di Matteo}, {Diakite}, {Diener},
  {Distefano}, {Dolding}, {Eappachen}, {Edvardsson}, {Enke}, {Esquej}, {Fabre},
  {Fabrizio}, {Faigler}, {Fedorets}, {Fernique}, {Fienga}, {Figueras},
  {Fouron}, {Fragkoudi}, {Fraile}, {Franke}, {Gai}, {Garabato},
  {Garcia-Gutierrez}, {Garc{\'\i}a-Torres}, {Garofalo}, {Gavras}, {Gerlach},
  {Geyer}, {Giacobbe}, {Gilmore}, {Girona}, {Giuffrida}, {Gomel}, {Gomez},
  {Gonzalez-Santamaria}, {Gonz{\'a}lez-Vidal}, {Granvik},
  {Guti{\'e}rrez-S{\'a}nchez}, {Guy}, {Hauser}, {Haywood}, {Helmi}, {Hidalgo},
  {Hilger}, {H{\l}adczuk}, {Hobbs}, {Holland}, {Huckle}, {Jasniewicz},
  {Jonker}, {Juaristi Campillo}, {Julbe}, {Karbevska}, {Kervella}, {Khanna},
  {Kochoska}, {Kontizas}, {Kordopatis}, {Korn}, {Kostrzewa-Rutkowska},
  {Kruszy{\'n}ska}, {Lambert}, {Lanza}, {Lasne}, {Le Campion}, {Le Fustec},
  {Lebreton}, {Lebzelter}, {Leccia}, {Leclerc}, {Lecoeur-Taibi}, {Liao},
  {Licata}, {Lindstr{\o}m}, {Lister}, {Livanou}, {Lobel}, {Madrero Pardo},
  {Managau}, {Mann}, {Marchant}, {Marconi}, {Marcos Santos}, {Marinoni},
  {Marocco}, {Marshall}, {Martin Polo}, {Mart{\'\i}n-Fleitas}, {Masip},
  {Massari}, {Mastrobuono-Battisti}, {Mazeh}, {McMillan}, {Messina},
  {Michalik}, {Millar}, {Mints}, {Molina}, {Molinaro}, {Moln{\'a}r},
  {Montegriffo}, {Mor}, {Morbidelli}, {Morel}, {Morris}, {Mulone}, {Munoz},
  {Muraveva}, {Murphy}, {Musella}, {Noval}, {Ord{\'e}novic}, {Orr{\`u}},
  {Osinde}, {Pagani}, {Pagano}, {Palaversa}, {Palicio}, {Panahi}, {Pawlak},
  {Pe{\~n}alosa Esteller}, {Penttil{\"a}}, {Piersimoni}, {Pineau}, {Plachy},
  {Plum}, {Poggio}, {Poretti}, {Poujoulet}, {Pr{\v{s}}a}, {Pulone}, {Racero},
  {Ragaini}, {Rainer}, {Raiteri}, {Rambaux}, {Ramos}, {Ramos-Lerate}, {Re
  Fiorentin}, {Regibo}, {Reyl{\'e}}, {Ripepi}, {Riva}, {Rixon}, {Robichon},
  {Robin}, {Roelens}, {Rohrbasser}, {Romero-G{\'o}mez}, {Rowell}, {Royer},
  {Rybicki}, {Sadowski}, {Sagrist{\`a} Sell{\'e}s}, {Sahlmann}, {Salgado},
  {Salguero}, {Samaras}, {Sanchez Gimenez}, {Sanna}, {Santove{\~n}a},
  {Sarasso}, {Schultheis}, {Sciacca}, {Segol}, {Segovia}, {S{\'e}gransan},
  {Semeux}, {Shahaf}, {Siddiqui}, {Siebert}, {Siltala}, {Slezak}, {Smart},
  {Solano}, {Solitro}, {Souami}, {Souchay}, {Spagna}, {Spoto}, {Steele},
  {Steidelm{\"u}ller}, {Stephenson}, {S{\"u}veges}, {Szabados}, {Szegedi-Elek},
  {Taris}, {Tauran}, {Taylor}, {Teixeira}, {Thuillot}, {Tonello}, {Torra},
  {Torra}, {Turon}, {Unger}, {Vaillant}, {van Dillen}, {Vanel}, {Vecchiato},
  {Viala}, {Vicente}, {Voutsinas}, {Weiler}, {Wevers}, {Wyrzykowski}, {Yoldas},
  {Yvard}, {Zhao}, {Zorec}, {Zucker}, {Zurbach}, \& {Zwitter}}]{GaiaEDR3}
{Gaia Collaboration}, {Brown}, A.~G.~A., {Vallenari}, A., {et~al.} 2021, \aap,
  649, A1

\bibitem[{{Green} {et~al.}(2018){Green}, {Schlafly}, {Finkbeiner}, {Rix},
  {Martin}, {Burgett}, {Draper}, {Flewelling}, {Hodapp}, {Kaiser}, {Kudritzki},
  {Magnier}, {Metcalfe}, {Tonry}, {Wainscoat}, \& {Waters}}]{3Ddust}
{Green}, G.~M., {Schlafly}, E.~F., {Finkbeiner}, D., {et~al.} 2018, \mnras,
  478, 651

\bibitem[{{Hellings} \& {Downs}(1983)}]{hd83}
{Hellings}, R.~W., \& {Downs}, G.~S. 1983, ApJL, 265, L39

\bibitem[{{Jennings} {et~al.}(2018){Jennings}, {Kaplan}, {Chatterjee},
  {Cordes}, \& {Deller}}]{JenningsEtAl:2018}
{Jennings}, R.~J., {Kaplan}, D.~L., {Chatterjee}, S., {Cordes}, J.~M., \&
  {Deller}, A.~T. 2018, \apj, 864, 26

\bibitem[{Jones {et~al.}(2001--)Jones, Oliphant, Peterson, {et~al.}}]{scipy}
Jones, E., Oliphant, T., Peterson, P., {et~al.} 2001--, {SciPy}: Open source
  scientific tools for {Python}, , , [Online; accessed 2016-08-24].
\newblock \url{http://www.scipy.org/}

\bibitem[{{Kaplan} {et~al.}(2016){Kaplan}, {Kupfer}, {Nice}, {Irrgang},
  {Heber}, {Arzoumanian}, {Beklen}, {Crowter}, {DeCesar}, {Demorest}, {Dolch},
  {Ellis}, {Ferdman}, {Ferrara}, {Fonseca}, {Gentile}, {Jones}, {Jones},
  {Kreuzer}, {Lam}, {Levin}, {Lorimer}, {Lynch}, {McLaughlin}, {Miller}, {Ng},
  {Pennucci}, {Prince}, {Ransom}, {Ray}, {Spiewak}, {Stairs}, {Stovall},
  {Swiggum}, \& {Zhu}}]{kkn+16}
{Kaplan}, D.~L., {Kupfer}, T., {Nice}, D.~J., {et~al.} 2016, \apj, 826, 86

\bibitem[{{Kopeikin}(1995)}]{Kopeikin}
{Kopeikin}, S.~M. 1995, \apjl, 439, L5

\bibitem[{{Lazaridis} {et~al.}(2009){Lazaridis}, {Wex}, {Jessner}, {Kramer},
  {Stappers}, {Janssen}, {Desvignes}, {Purver}, {Cognard}, {Theureau}, {Lyne},
  {Jordan}, \& {Zensus}}]{lwj+09}
{Lazaridis}, K., {Wex}, N., {Jessner}, A., {et~al.} 2009, \mnras, 400, 805

\bibitem[{{Li} {et~al.}(2018){Li}, {Yang}, {An}, {Paragi}, {Deller},
  {Reynolds}, {Hong}, {Wang}, {Ding}, {Xia}, {Yan}, \& {Guo}}]{lyp+18}
{Li}, Z., {Yang}, J., {An}, T., {et~al.} 2018, \mnras, 476, 399

\bibitem[{{Lindegren} {et~al.}(2018){Lindegren}, {Hernandez}, {Bombrun},
  {Klioner}, {Bastian}, {Ramos-Lerate}, {de Torres}, {Steidelmuller},
  {Stephenson}, {Hobbs}, {Lammers}, {Biermann}, {Geyer}, {Hilger}, {Michalik},
  {Stampa}, {McMillan}, {Castaneda}, {Clotet}, {Comoretto}, {Davidson},
  {Fabricius}, {Gracia}, {Hambly}, {Hutton}, {Mora}, {Portell}, {van Leeuwen},
  {Abbas}, {Abreu}, {Altmann}, {Andrei}, {Anglada}, {Balaguer-Nunez},
  {Barache}, {Becciani}, {Bertone}, {Bianchi}, {Bouquillon}, {Bourda},
  {Brusemeister}, {Bucciarelli}, {Busonero}, {Buzzi}, {Cancelliere},
  {Carlucci}, {Charlot}, {Cheek}, {Crosta}, {Crowley}, {de Bruijne}, {de
  Felice}, {Drimmel}, {Esquej}, {Fienga}, {Fraile}, {Gai}, {Garralda},
  {Gonzalez-Vidal}, {Guerra}, {Hauser}, {Hofmann}, {Holl}, {Jordan},
  {Lattanzi}, {Lenhardt}, {Liao}, {Licata}, {Lister}, {Loffler}, {Marchant},
  {Martin-Fleitas}, {Messineo}, {Mignard}, {Morbidelli}, {Poggio}, {Riva},
  {Rowell}, {Salguero}, {Sarasso}, {Sciacca}, {Siddiqui}, {Smart}, {Spagna},
  {Steele}, {Taris}, {Torra}, {van Elteren}, {van Reeven}, \&
  {Vecchiato}}]{lhb+18}
{Lindegren}, L., {Hernandez}, J., {Bombrun}, A., {et~al.} 2018, ArXiv e-prints,
  arXiv:1804.09366

\bibitem[{{Lorimer} \& {Kramer}(2004)}]{lk04}
{Lorimer}, D.~R., \& {Kramer}, M. 2004, {Handbook of Pulsar Astronomy}
  (Cambridge University Press)

\bibitem[{{Luri} {et~al.}(2018){Luri}, {Brown}, {Sarro}, {Arenou},
  {Bailer-Jones}, {Castro-Ginard}, {de Bruijne}, {Prusti}, {Babusiaux}, \&
  {Delgado}}]{lbs18}
{Luri}, X., {Brown}, A.~G.~A., {Sarro}, L.~M., {et~al.} 2018, \aap, 616, A9

\bibitem[{{Lutz} \& {Kelker}(1973)}]{lk73}
{Lutz}, T.~E., \& {Kelker}, D.~H. 1973, \pasp, 85, 573

\bibitem[{{Lyne} \& {Graham-Smith}(2012)}]{pulsarAstronomy}
{Lyne}, A., \& {Graham-Smith}, F. 2012, {Pulsar Astronomy} (Cambridge
  University Press)

\bibitem[{{Manchester} {et~al.}(2013){Manchester}, {Hobbs}, {Bailes}, {Coles},
  {van Straten}, {Keith}, {Shannon}, {Bhat}, {Brown}, {Burke-Spolaor},
  {Champion}, {Chaudhary}, {Edwards}, {Hampson}, {Hotan}, {Jameson}, {Jenet},
  {Kesteven}, {Khoo}, {Kocz}, {Maciesiak}, {Oslowski}, {Ravi}, {Reynolds},
  {Sarkissian}, {Verbiest}, {Wen}, {Wilson}, {Yardley}, {Yan}, \&
  {You}}]{mhb+13}
{Manchester}, R.~N., {Hobbs}, G., {Bailes}, M., {et~al.} 2013, Publications of
  the Astronomical Society of Australia, 30, 17

\bibitem[{{Mingarelli}(2019)}]{M19}
{Mingarelli}, C. M.~F. 2019, Nature Astronomy, 3, 8

\bibitem[{{Mingarelli} {et~al.}(2012){Mingarelli}, {Grover}, {Sidery}, {Smith},
  \& {Vecchio}}]{mgs+12}
{Mingarelli}, C.~M.~F., {Grover}, K., {Sidery}, T., {Smith}, R.~J.~E., \&
  {Vecchio}, A. 2012, Physical Review Letters, 109, 081104

\bibitem[{{Mingarelli} {et~al.}(2017){Mingarelli}, {Lazio}, {Sesana}, {Greene},
  {Ma}, {Ellis}, {Croft}, {Burke-Spolaor}, \& {Taylor}}]{mls+17}
{Mingarelli}, C.~M.~F., {Lazio}, T.~J.~W., {Sesana}, A., {et~al.} 2017, Nature
  Astronomy, 1, 886

\bibitem[{{Nelson} {et~al.}(2004){Nelson}, {Dubeau}, \& {MacCannell}}]{ndm+04}
{Nelson}, L.~A., {Dubeau}, E., \& {MacCannell}, K.~A. 2004, \apj, 616, 1124

\bibitem[{{Nicastro} {et~al.}(1995){Nicastro}, {Lyne}, {Lorimer}, {Harrison},
  {Bailes}, \& {Skidmore}}]{n95}
{Nicastro}, L., {Lyne}, A.~G., {Lorimer}, D.~R., {et~al.} 1995, \mnras, 273,
  L68

\bibitem[{{Perera} {et~al.}(2019){Perera}, {DeCesar}, {Demorest}, {Kerr},
  {Lentati}, {Nice}, {Oslowski}, {Ransom}, {Keith}, {Arzoumanian}, {Bailes},
  {Baker}, {Bassa}, {Bhat}, {Brazier}, {Burgay}, {Burke-Spolaor}, {Caballero},
  {Champion}, {Chatterjee}, {Chen}, {Cognard}, {Cordes}, {Crowter}, {Dai},
  {Desvignes}, {Dolch}, {Ferdman}, {Ferrara}, {Fonseca}, {Goldstein},
  {Graikou}, {Guillemot}, {Hazboun}, {Hobbs}, {Hu}, {Islo}, {Janssen},
  {Karuppusamy}, {Kramer}, {Lam}, {Lee}, {Liu}, {Luo}, {Lyne}, {Manchester},
  {McKee}, {McLaughlin}, {Mingarelli}, {Parthasarathy}, {Pennucci}, {Perrodin},
  {Possenti}, {Reardon}, {Russell}, {Sanidas}, {Sesana}, {Shaifullah},
  {Shannon}, {Siemens}, {Simon}, {Spiewak}, {Stairs}, {Stappers}, {Swiggum},
  {Taylor}, {Theureau}, {Tiburzi}, {Vallisneri}, {Vecchio}, {Wang}, {Zhang},
  {Zhang}, {Zhu}, \& {Zhu}}]{iptadr2}
{Perera}, B.~B.~P., {DeCesar}, M.~E., {Demorest}, P.~B., {et~al.} 2019, arXiv
  e-prints, arXiv:1909.04534

\bibitem[{{Phinney}(2001)}]{p01}
{Phinney}, E.~S. 2001, ArXiv Astrophysics e-prints, astro-ph/0108028

\bibitem[{{Porayko} {et~al.}(2018){Porayko}, {Zhu}, {Levin}, {Hui}, {Hobbs},
  {Grudskaya}, {Postnov}, {Bailes}, {Bhat}, {Coles}, {Dai}, {Dempsey}, {Keith},
  {Kerr}, {Kramer}, {Lasky}, {Manchester}, {Os{\l}owski}, {Parthasarathy},
  {Ravi}, {Reardon}, {Rosado}, {Russell}, {Shannon}, {Spiewak}, {van Straten},
  {Toomey}, {Wang}, {Wen}, {You}, \& {PPTA Collaboration}}]{pzl+18}
{Porayko}, N.~K., {Zhu}, X., {Levin}, Y., {et~al.} 2018, \prd, 98, 102002

\bibitem[{{Reardon} {et~al.}(2016){Reardon}, {Hobbs}, {Coles}, {Levin},
  {Keith}, {Bailes}, {Bhat}, {Burke-Spolaor}, {Dai}, {Kerr}, {Lasky},
  {Manchester}, {Os{\l}owski}, {Ravi}, {Shannon}, {van Straten}, {Toomey},
  {Wang}, {Wen}, {You}, \& {Zhu}}]{rhc+16}
{Reardon}, D.~J., {Hobbs}, G., {Coles}, W., {et~al.} 2016, \mnras, 455, 1751

\bibitem[{{Sazhin}(1978)}]{saz78}
{Sazhin}, M.~V. 1978, \sovast, 22, 36

\bibitem[{{Schlegel} {et~al.}(1998){Schlegel}, {Finkbeiner}, \& {Davis}}]{sfd}
{Schlegel}, D.~J., {Finkbeiner}, D.~P., \& {Davis}, M. 1998, \apj, 500, 525

\bibitem[{{Shklovskii}(1970)}]{s70}
{Shklovskii}, I.~S. 1970, \sovast, 13, 562

\bibitem[{{Smits} {et~al.}(2011){Smits}, {Tingay}, {Wex}, {Kramer}, \&
  {Stappers}}]{stw+11}
{Smits}, R., {Tingay}, S.~J., {Wex}, N., {Kramer}, M., \& {Stappers}, B. 2011,
  \aap, 528, A108

\bibitem[{{Stairs} {et~al.}(2005){Stairs}, {Faulkner}, {Lyne}, {Kramer},
  {Lorimer}, {McLaughlin}, {Manchester}, {Hobbs}, {Camilo}, {Possenti},
  {Burgay}, {D'Amico}, {Freire}, \& {Gregory}}]{sfl+05}
{Stairs}, I.~H., {Faulkner}, A.~J., {Lyne}, A.~G., {et~al.} 2005, \apj, 632,
  1060

\bibitem[{{Tauris} \& {van den Heuvel}(2006)}]{tvdh06}
{Tauris}, T.~M., \& {van den Heuvel}, E.~P.~J. 2006, {Formation and evolution
  of compact stellar X-ray sources} (Cambridge University Press), 623--665

\bibitem[{{The Astropy Collaboration} {et~al.}(2013){The Astropy
  Collaboration}, {Robitaille, Thomas P.}, {Tollerud, Erik J.}, {Greenfield,
  Perry}, {Droettboom, Michael}, {Bray, Erik}, {Aldcroft, Tom}, {Davis, Matt},
  {Ginsburg, Adam}, {Price-Whelan, Adrian M.}, {Kerzendorf, Wolfgang E.},
  {Conley, Alexander}, {Crighton, Neil}, {Barbary, Kyle}, {Muna, Demitri},
  {Ferguson, Henry}, {Grollier, FrÃÂ©dÃÂ©ric}, {Parikh, Madhura M.},
  {Nair, Prasanth H.}, {GÃÅnther, Hans M.}, {Deil, Christoph}, {Woillez,
  Julien}, {Conseil, Simon}, {Kramer, Roban}, {Turner, James E. H.}, {Singer,
  Leo}, {Fox, Ryan}, {Weaver, Benjamin A.}, {Zabalza, Victor}, {Edwards,
  Zachary I.}, {Azalee Bostroem, K.}, {Burke, D. J.}, {Casey, Andrew R.},
  {Crawford, Steven M.}, {Dencheva, Nadia}, {Ely, Justin}, {Jenness, Tim},
  {Labrie, Kathleen}, {Lim, Pey Lian}, {Pierfederici, Francesco}, {Pontzen,
  Andrew}, {Ptak, Andy}, {Refsdal, Brian}, {Servillat, Mathieu}, \& {Streicher,
  Ole}}]{astropy}
{The Astropy Collaboration}, {Robitaille, Thomas P.}, {Tollerud, Erik J.},
  {et~al.} 2013, A\&A, 558, A33.
\newblock \url{http://dx.doi.org/10.1051/0004-6361/201322068}

\bibitem[{{The WFIRST Astrometry Working Group} {et~al.}(2017){The WFIRST
  Astrometry Working Group}, {Sanderson}, {Bellini}, {Casertano}, {Lu},
  {Melchior}, {Bennett}, {Shao}, {Rhodes}, {Malhotra}, {Gaudi}, {Fall},
  {Nelan}, {Guhathakurta}, {Anderson}, {Ho}, \& {Libralato}}]{WFIRST}
{The WFIRST Astrometry Working Group}, {Sanderson}, R.~E., {Bellini}, A.,
  {et~al.} 2017, arXiv e-prints, arXiv:1712.05420

\bibitem[{{van den Heuvel}(1984)}]{vdh84}
{van den Heuvel}, E.~P.~J. 1984, Journal of Astrophysics and Astronomy, 5, 209

\bibitem[{van~der Walt {et~al.}(2011)van~der Walt, Colbert, \&
  Varoquaux}]{numpy}
van~der Walt, S., Colbert, S.~C., \& Varoquaux, G. 2011, Computing in Science
  Engineering, 13, 22

\bibitem[{{Verbiest} {et~al.}(2008){Verbiest}, {Bailes}, {van Straten},
  {Hobbs}, {Edwards}, {Manchester}, {Bhat}, {Sarkissian}, {Jacoby}, \&
  {Kulkarni}}]{vbs+08}
{Verbiest}, J.~P.~W., {Bailes}, M., {van Straten}, W., {et~al.} 2008, ApJ, 679,
  675

\bibitem[{{Xin} {et~al.}(2021){Xin}, {Mingarelli}, \& {Hazboun}}]{Xin2021}
{Xin}, C., {Mingarelli}, C. M.~F., \& {Hazboun}, J.~S. 2021, \apj, 915, 97

\bibitem[{{Yao} {et~al.}(2017){Yao}, {Manchester}, \& {Wang}}]{ymw+17}
{Yao}, J.~M., {Manchester}, R.~N., \& {Wang}, N. 2017, \apj, 835, 29

\bibitem[{{Zhu} {et~al.}(2016){Zhu}, {Wen}, {Xiong}, {Xu}, {Wang}, {Mohanty},
  {Hobbs}, \& {Manchester}}]{zwx+16}
{Zhu}, X.-J., {Wen}, L., {Xiong}, J., {et~al.} 2016, \mnras, 461, 1317

\end{thebibliography}
\end{document}